# Learning by Collaborating in the Age of AI:
# Research Mentorship for Polymathic, Workflow-Centered Science

Tian Zheng

Department of Statistics, Columbia University, New York, NY, USA

Corresponding author: Tian Zheng

Department of Statistics, Columbia University,

1255 Amsterdam Avenue, New York, NY 10027, USA

Email: tz33@columbia.edu | Tel: 212-851-2149

**Abstract**

Artificial intelligence (AI) is changing how data-driven research is produced. As AI tools make information and technical execution more accessible, mentorship must concentrate on judgment, ethics, and the design of end-to-end AI-augmented workflows. Disciplinary experts need mentoring to adopt AI while preserving their expertise and human agency, and AI-native research trainees need mentoring to become polymathic workflow developers. This chapter proposes *Everyone's Backyard*, an interdisciplinary model in which disciplinary experts, trainees, and community partners mentor one another in both directions and across disciplines through shared research practice in a *convergence studio*. The model centers on four capacities: statistical thinking, data ethics, collaboration, and AI for social good and society. We illustrate the model with the Spring 2026 aiX Faculty Fellows Program at Columbia University, which brought together 20 faculty fellows and 15 student interns from 14 schools and departments. Evaluation data show growth in faculty confidence, intern competence, and cross-disciplinary connection.

**Keywords**: research mentorship; AI-augmented research; peer mentoring; reverse mentoring; polymathic training; workflow-centered science

## I. Introduction

Artificial intelligence (AI) is changing how data-driven research is conducted and how research capacities are developed. Traditional mentorship has centered on disciplinary apprenticeship: a trainee learns the norms, methods, and standards of a field from an established expert. That relationship remains essential. At the same time, AI tools can now support many activities that once required close supervision, including literature search, coding, data cleaning, visualization, drafting, and dissemination [1, 2]. A trainee can now obtain definitions, examples, and basic technical guidance independently. The mentor's contribution therefore moves toward the decisions that shape a research project from beginning to end.

Those decisions include identifying a worthwhile question, determining what evidence is sufficient, examining how data were generated, recognizing when an AI-produced result is wrong, and interpreting a finding in context. Early evidence also suggests that uncritical reliance on AI can weaken the reasoning and ownership that research training is meant to develop [3]. Research mentorship from experts to trainees becomes more important as these forms of judgment become central to responsible inquiry. They are developed through practice, feedback, and participation in real work.

The shift to AI-augmented research also creates a parallel challenge for disciplinary experts, including faculty, senior scientists, and other established researchers and practitioners. Many are eager to engage AI while often lacking a clear starting point. Limited AI literacy makes it difficult to identify needed support, while the absence of structured support slows the development of that literacy. A Spring 2026 survey of 26 department chairs and division heads at Columbia University illustrates the institutional urgency. At least 20 respondents rated each of three framings of AI as moderately to extremely urgent for their unit: AI as academic work, student use of AI in courses, and AI for social good and society (Figure 1).

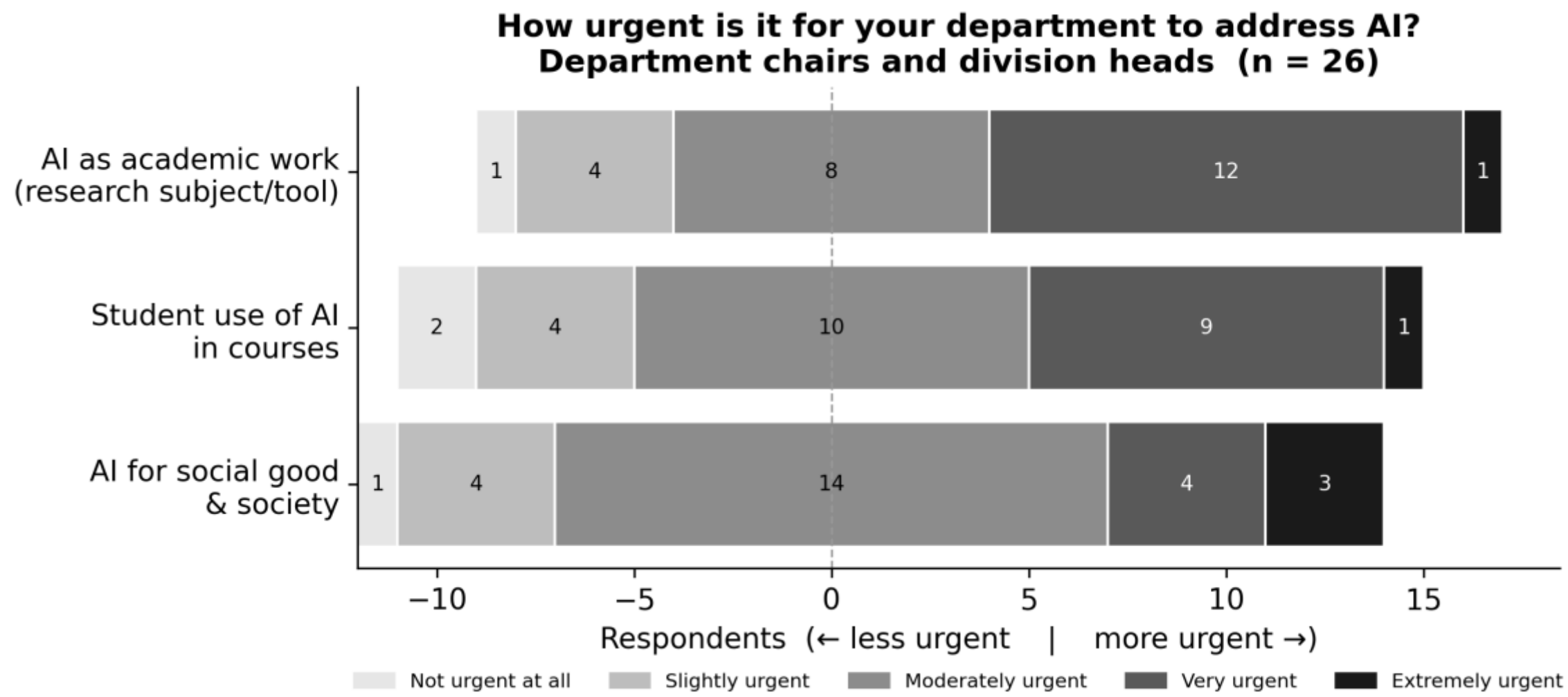

*Figure 1. Department leaders' ratings of the urgency of addressing AI under three framings. The complete-case sample includes 26 respondents. AI as academic work receives the strongest ratings, while AI for social good and society is concentrated at moderate urgency.*

The pace of AI development compounds this challenge. Each new wave of tools requires disciplines to revisit practices, curricula, and standards. The response time of a discipline affects how quickly it can recover, learn, and build on the next opportunity; the response-time argument is developed in Section II.B and illustrated in Figure 2.

Two implications follow for how research capacities should be developed. First, because AI now reaches nearly every stage of a project, the stages can no longer be mastered in isolation. A choice made during data cleaning or prompting shapes what can be claimed at the end, and an error introduced early can pass unnoticed through later steps. The workflow as a whole, rather than any single technique, becomes what a researcher must learn to design and evaluate.

Second, research mentoring is now needed in both directions. Disciplinary experts need research mentoring to bring AI into their work while preserving their expertise and human agency. AI-native research trainees, such as students who are often more familiar with AI tools than their mentors, need mentoring to grow from tool users into polymathic workflow developers who can frame a question, judge evidence, and carry a project from problem to use [4]. Each group holds part of what the other lacks. Disciplinary experts bring depth and judgment but are often still building AI literacy; trainees bring fluency with the tools but are still developing that judgment; community partners bring knowledge of how data are generated and what a finding means in context. Because the expertise that AI-enabled workflows demand rarely resides in one person, one-directional apprenticeship from expert to trainee cannot meet these needs. Mentoring must become bidirectional between experts and trainees, and multidirectional among experts and trainees through peer mentoring. Collaboration becomes the setting in which these forms of expertise meet and check one another, not an add-on to training.

Mentorship in data-driven research should therefore train scholars to learn by collaborating within AI-enabled, end-to-end workflows. This chapter proposes *Everyone's Backyard*, a model of research mentorship built around an interdisciplinary *convergence studio*. The name draws on John Tukey's observation that statisticians get to "play in everyone's backyard" and his call to study how data analysis is practiced across fields [5]. In the model, a local problem becomes a point of entry into shared inquiry, and every participant both mentors and is mentored. Each person brings a discipline into the shared backyard that everyone is invited into: disciplinary experts, trainees, and community partners lead where their expertise is deepest and learn from others where it is not. Working through the workflow together, they develop knowledge that can travel across settings. Recent work argues that research mentorship is shifting from apprenticeship toward partnership with AI [6]; Everyone's Backyard extends that shift by treating mentoring as a partnership among people as well as with AI.

The *Everyone's Backyard* model has five design elements. Bidirectional and multidirectional mentoring matches each need to a source of expertise: disciplinary experts receive reverse mentoring from trainees on tools and emerging practice [7], trainees receive project-based mentoring from experts on disciplinary judgment and research design, and both groups receive peer mentoring across disciplines through a community of practice and learning community [8-11]. Workflow-centered science makes the complete arc of inquiry the unit of training. The convergence studio provides a shared setting for co-design and co-building around real problems. A community of practice sustains the work through a regular rhythm, multiple levels of participation, and a shared repertoire of tools and lessons. Four capacities orient the work: statistical thinking, data ethics, collaboration, and AI for social good and society.

The design also addresses two risks. The first is shallowness: training across fields can trade depth for breadth. The model guards against this by grounding each participant in a deep disciplinary project and building the ability to connect that home to other fields. The second is unequal access: AI can widen gaps between disciplines that have resources for tools and expertise and those that do not. Shared infrastructure and inclusive studios build capacity in both.

Section II describes the model. Section III presents the Spring 2026 aiX Faculty Fellows Program at Columbia University as a case study of its implementation, and Section IV summarizes implications for research mentorship in the age of AI.

## II. The Everyone's Backyard Model

### A. Teaching End-to-End Research Workflows

AI changes the balance between technical execution and research design. A trainee may learn how to run a model, clean a dataset, or draft a summary with assistance from an AI system. The mentor's work is to help the trainee understand how these activities connect and where each decision affects the integrity of the whole inquiry. The complete workflow includes problem formulation and stakeholder engagement; evidence synthesis; data acquisition and curation; study design and statistical reasoning; AI-assisted analysis; validation and interpretation; communication and implementation; and reflection.

Mentors and trainees can treat the workflow as a design problem. They begin by understanding the problem and the people it affects, decompose the inquiry into stages, specify where AI can assist, identify points that require human review, and revise the workflow as evidence accumulates. The mentor serves as a workflow architect who helps the trainee see research as a connected system. This emphasis builds on Tukey's attention to the practice of data analysis and Donoho's account of reproducible, cross-workflow data science [5, 12].

The workflow should make its decision points visible. Human judgment remains responsible for problem selection, assumptions, source verification, interpretation, and decisions about use. Trainees verify AI-cited sources, inspect the data-generating process, state assumptions, and characterize remaining uncertainty before relying on an output. They document where, how, and why AI was used, drawing on model cards and model information sheets as practical examples [13, 14]. They design for predictability, computability, and stability [15], and ask who is represented in the data and who will be affected by the conclusions.

A workflow can also be organized into reusable blocks. A reproducible data pipeline, bias-audit routine, validation protocol, or dissemination template can be adapted from one project to another. Reuse lowers the cost of entry for a new field and creates a concrete reason for collaborators to share practices. The resulting training is modular while remaining grounded in context. As agentic AI systems begin to orchestrate multistep processes, mentors must help trainees evaluate the design and oversight of the workflow as a whole.

### B. Collaboration and the Convergence Studio

The convergence studio is the structural metaphor and practical setting for Everyone's Backyard. It is a physical or virtual space where disciplinary experts, trainees, statisticians, and community partners work around real problems. The model draws on *communities of practice*, in which newcomers develop expertise through participation and a group builds a shared repertoire of tools, language, and practices [8-10]. It also draws on *faculty learning communities*, which connect otherwise isolated faculty and sustain collaborative work around shared questions [11].

Convergence develops through shared learning, shared tools, and sustained collaboration. Participants ask naive questions, translate disciplinary language, and identify overlapping problems. AI platforms, coding environments, data infrastructure, and statistical software create a common technical language. Project teams then co-own problems, co-design workflows, and co-author knowledge over time. Co-design research describes this process as a form of learning in which teams bring multiple kinds of expertise to a shared task [16].

Communication and translation are central studio practices. Disciplinary experts contribute purpose and learning objectives. Trainees translate those objectives into technical designs, while statisticians help teams formulate questions and evaluate evidence. Collaboration can range from a Translator pattern, in which trainees turn experts' goals into technical designs, to a more equal Co-Pilot partnership. Trainees often mentor experts on tools while learning disciplinary context in return. Experts also mentor one another across disciplines through tool demonstrations and shared troubleshooting. This reciprocal pattern extends research on collaborative, network-based, and systems mentorship, mentor benefits, learning by teaching, and *reverse mentoring* [7, 17-23].

A community of practice gives the studio continuity. Regular cohort meetings establish a rhythm; disciplinary experts, near-peer mentors, and trainees provide different levels of participation; and documented tools and workflows form a shared repertoire that future cohorts can extend [10].

The studio addresses agility and access as well as collaboration. The Satir change model describes how a disruptive innovation can produce an initial decline in confidence and productivity before a group reaches a new equilibrium [24-26]. When disruptions recur, response time determines whether a field can recover and build iteratively. Figure 2 adapts this model to repeated waves of AI development. Shared infrastructure, consultation, and near-peer support shorten the time required to move from disruption to productive experimentation. Support calibrated to participants' different starting points in technical confidence shortens this time further.

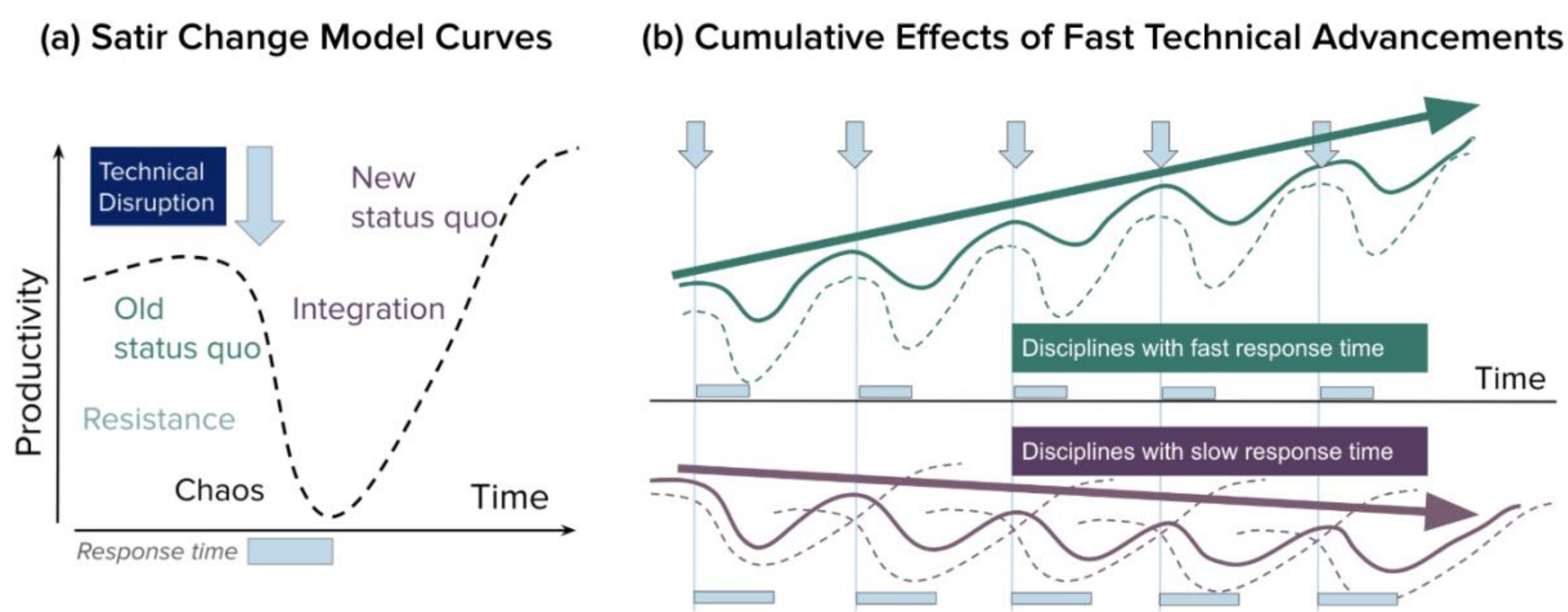

*Figure 2. The Satir change model applied to repeated waves of AI disruption. Panel (a) shows an initial decline followed by integration and a new equilibrium. Panel (b) shows how shorter response times allow disciplines to accumulate gains across repeated disruptions. Adapted from the Satir change model as discussed by Cheung [24] and Saunders, Coulby, and Lindsay [25].*

A studio requires institutional support. Dedicated trainee support, clear policy guidance, funding for computation, and formal recognition for participation are conditions for continuity. These conditions are part of the mentoring model because they determine whether collaborative research can continue long enough for shared judgment and shared practice to develop.

### C. Statistical Thinking as a Shared Research Capacity

Statistical thinking is a core component of AI literacy [27]. AI systems learn from data, produce probabilistic outputs, and generalize under uncertainty. Responsible use therefore requires the ability to formulate questions, evaluate evidence, interpret uncertainty, and make decisions in context. AI can generate analyses, summaries, visualizations, and models. Trainees still need to ask what assumptions are being made, how the data were generated, what uncertainty remains, what biases may be present, and which claims the evidence supports.

In the convergence studio, statistical thinking connects domain questions to rigorous evidence. The statistician studio lead's role includes teaching concepts and modeling habits of inquiry: checking the estimand, examining measurement, distinguishing association from causation, and communicating uncertainty to collaborators and decision-makers. This capacity belongs to every participant, with depth calibrated to the role each person plays in the workflow. It gives interdisciplinary teams a shared basis for evaluating AI-assisted results [12, 15].

### D. Data Ethics: Bias, Privacy, and Responsible AI Use

Data ethics is woven through the workflow. Bias can enter during data collection, labeling, model training, and interpretation, and AI can amplify inequities that are already present. Projects that audit a widely used risk-assessment tool or examine representation in a training dataset make ethical questions part of the work that trainees are learning to do.

Privacy requires attention to consent, governance, retention, access controls, de-identification limits, and the tension between data utility and individual rights. Workflows can build protections in from the start, such as blurring faces in images, stripping metadata, and avoiding centralized storage of sensitive data.

Responsible AI use also requires documentation, validation, reproducibility, and attention to automation bias. When institutional guidance is unsettled, as with the use of copyrighted material in AI tools, the community can surface the question, work with institutional offices to obtain guidance, and share what it learns across projects. Equity completes the framework: AI-enabled workflows should expand participation and capacity across institutions rather than concentrate advantages where resources are already abundant. These practices complement broader guidance on responsible AI research, upstream bias, and inclusive access [28-30].

Mentors cultivate ethical judgment by embedding these questions at each stage of practice. Ethics becomes part of how a question is formulated, a dataset is assembled, a model is evaluated, and a result is communicated. Participants learn to explain the reasons behind their decisions and to revise a workflow when the evidence or the affected community requires it.

### E. AI for Social Good and Society

AI for social good and society supplies a shared purpose for the convergence studio. The studio convenes around problems that matter to communities, public institutions, and professional practice. Its social orientation has three dimensions. Problem sensing and domain curiosity help researchers notice questions worth studying in contexts beyond their own. Societal orientation directs inquiry toward public health, education, environmental sustainability, justice, and community well-being. Impact and accountability require researchers to consider who benefits, who bears risk, and whether a tool should be built alongside whether it can be built [31-33].

The mentor cultivates this orientation by helping trainees see local problems as opportunities for socially useful inquiry. Community and domain partners can clarify needs, identify unacceptable risks, and evaluate whether a workflow produces value in practice. This approach connects the Everyone's Backyard model to community-engaged AI research and to the broader case for interdisciplinary partnerships that direct AI toward genuine societal benefit [31-33].

### F. Learning by Collaborating and Polymathic Training

The capacity for workflow-centered science develops through collaborative practice. Trainees work on projects that include ambiguity, imperfect data, conflicting interpretations, and ethical tension. They rotate among roles such as problem framer, data steward, statistical analyst, AI auditor, ethics reviewer, domain translator, and dissemination lead. The work develops integrative depth: a strong disciplinary home combined with working fluency across statistical reasoning, computational and AI literacy, ethics, communication, implementation, and workflow design [4].

This form of polymathic training is supported by project-based mentorship teams, workflow audits, AI-use documentation, reproducibility checks, reflective research memos, and community- or domain-partnered projects. It also depends on reciprocal learning. Disciplinary experts bring depth and judgment; trainees and near-peer mentors often bring experience with tools, interfaces, and emerging practices. Explaining a method, translating a concept, or testing a prototype can deepen the mentor's own understanding [20].

## III. Case Study: The Spring 2026 aiX Faculty Fellows Program

The Spring 2026 aiX Faculty Fellows Program served as a pilot of the Everyone's Backyard model. The program brought together 20 faculty fellows and 15 student interns from 14 schools and departments at Columbia University. Over one semester, the cohort developed 17 projects across disciplinary clusters, including law, health sciences, language education, social science and policy, business, psychology, and STEM education (Figure 3) [34, 35]. The program tested the conditions for a convergence studio in the context of faculty-led AI education, which offers a useful setting for examining research mentorship.

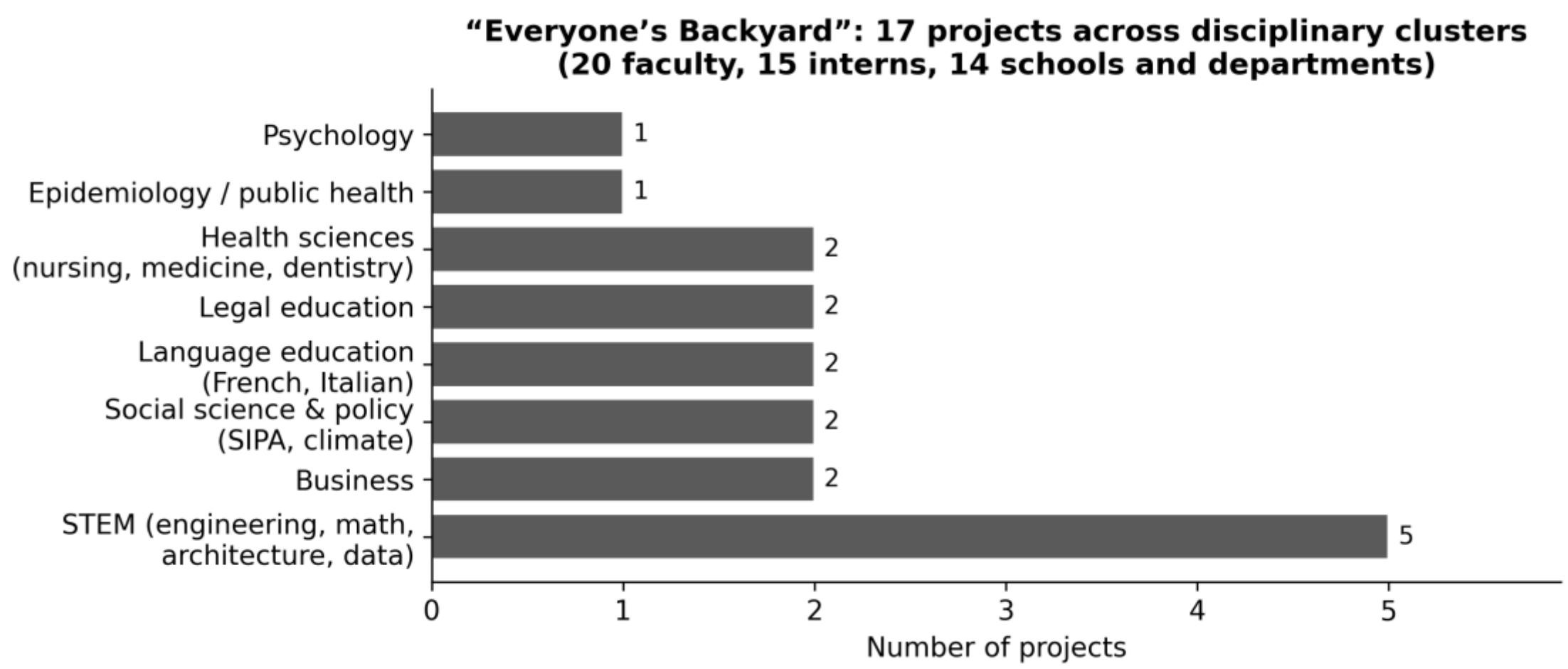

*Figure 3. Everyone's Backyard in practice. The Spring 2026 aiX Faculty Fellows Program developed 17 projects across disciplinary clusters with 20 faculty fellows, 15 student interns, and 14 schools and departments.*

The program ran from January through early May 2026 and combined a full-day bootcamp, bi-weekly cohort meetings, weekly intern studio hours, individualized faculty support, and a public showcase. Evidence for this chapter comes from the program's end-of-semester faculty and intern reviews, weekly research reflections, project status surveys, and the department-leader survey. The analysis is descriptive. The participant samples were self-selected, faculty reviews were anonymous and not matched to baseline responses, and the weekly reflections are self-reported. The pilot supports formative interpretation rather than causal inference.

### Example 1: The Convergence Studio in Action

The aiX program made the studio operational. The January bootcamp introduced AI concepts, institutional resources, and design frameworks. Extended introductions, in which each fellow described a project vision, were later identified as the most valuable part of the day. Bi-weekly cohort meetings provided peer-led tool demonstrations, pedagogical discussion, and troubleshooting. Weekly studio hours with interns supported project updates and technical guidance while respecting faculty time. The semester ended with a public showcase at which 13 teams presented their work [34, 35].

Within teams, collaboration followed the reciprocal pattern described in Section II.B. The dominant pattern was a Translator model, and several pairs developed a more equal Co-Pilot partnership. Interns often mentored faculty on tools while learning disciplinary context in return, and faculty fellows mentored one another across disciplines through peer-led tool demonstrations and shared troubleshooting in cohort meetings. Faculty rated the fellowship's effect on their professional network and community positively, and they rated the bi-weekly cohort meetings as a meaningful source of connection (Figure 4).

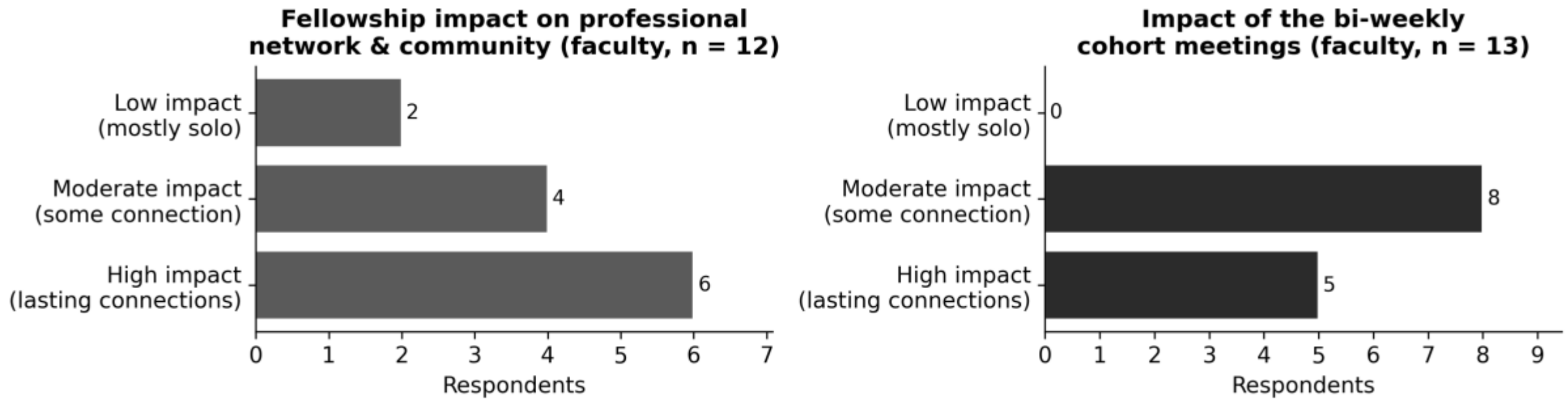

*Figure 4. The studio as a community of practice. Faculty reported the fellowship's impact on their professional network and community (n = 12) and the impact of bi-weekly cohort meetings (n = 13).*

The weekly reflections show how this rhythm supported project work. Across 105 reflections from 15 program weeks, interns reported progress on a 1 to 5 scale and indicated whether they had discussed progress with the faculty lead (Figure 5). The two measures dip together in Week 8 and recover later in the semester. This descriptive association is consistent with the value of regular mentoring contact, while the data do not establish a causal relationship. The pattern also shows why workflow-centered mentoring needs recurring touchpoints rather than a single training event.

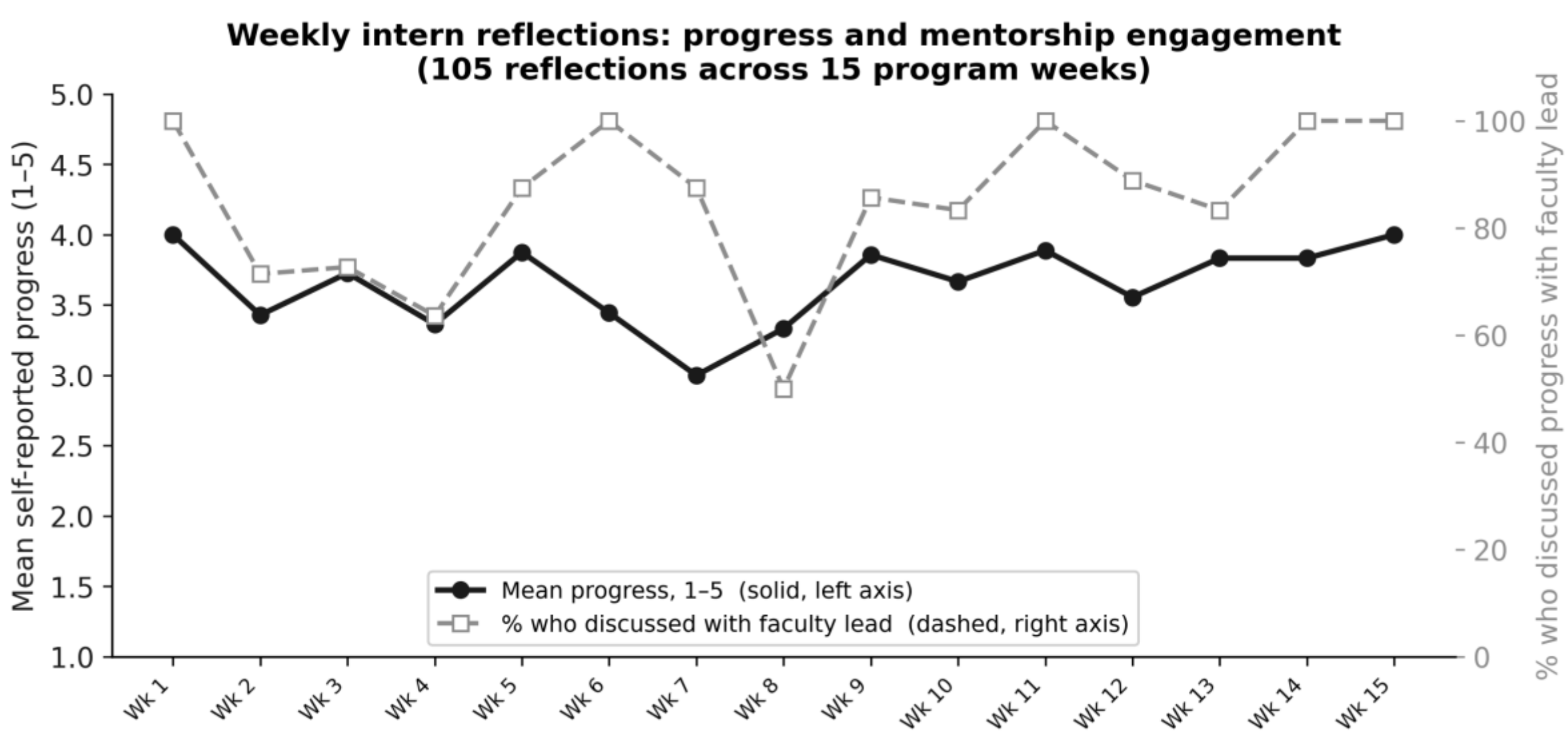

*Figure 5. Weekly intern reflections across 15 program weeks (105 reflections). The solid line shows mean self-reported progress on a 1 to 5 scale, and the dashed line shows the percentage of reflections reporting discussion with the faculty lead. Both measures dip in Week 8; the association is descriptive.*

Throughout the semester, the mentor guided project workflows with questions about formulation, assumptions, uncertainty, evidence, and audience. The questions made the mentor's role visible: the mentor did not need to supply every technical step, but did need to help the team decide what the step was for, how it should be checked, and what would count as a useful result.

### Example 2: Designing for Evidence and Explanation

Faculty fellows from radiology and dental medicine collaborated on a problem their disciplines share: as AI-augmented image interpretation becomes the norm, students must learn to use AI tools rigorously while still developing their own diagnostic expertise. A model that simply labels an image invites students to accept its answer. The design task was instead to create a learning experience in which students test that answer against the evidence. For the interns, the project showed that designing such an experience requires applying explainable machine learning methods to computer vision models, for example by showing which regions of an image drive a model's prediction. Students can then compare the model's evidence with their own reading of the image, ask where and why the two diverge, and practice the disciplinary judgment needed to accept, question, or override an AI result.

The same evidence- and explanation-based approach shaped projects that designed AI discussion partners. Rather than supplying answers, the AI asked follow-up questions based on each student's response, inviting the student to return to the source materials, weigh the evidence, and justify a position. In both cases, the discipline supplied the problem, while statistical thinking and data ethics shaped the workflow for addressing it. The shared design principle was AI that makes reasoning visible and keeps students' critical thinking, and their growing disciplinary expertise, at the center of the work.

The aiX projects also show how the social orientation described in Section II.E can begin in a local setting. A nursing team built a mobile-first virtual instructor for students on international clinical placements, addressing the loss of on-site mentorship that students often experience abroad. A chemical-engineering team developed a simulated public hearing in which students respond to AI-generated community stakeholders with conflicting concerns. These projects connect technical design to professional judgment, public participation, and the people who will encounter the resulting tools.

### Example 3: Role Rotation and the Collaborative Workflow

The intern cohort shows how role rotation builds polymathic capacity. Students from statistics contributed methodological rigor, students from education brought learning theory, students from design contributed interface expertise, and students from computer science and engineering contributed technical build capacity. Across projects, interns moved among problem framing, data stewardship, technical building, AI auditing, ethics review, domain translation, and dissemination.

Ethics review was built into project work. A law project used the COMPAS recidivism tool [36, 37] to show how technical choices shape fairness outcomes, and an urban-planning project used a student-built image dataset to examine representation across age, gender expression, and skin tone. Another project used a bring-your-own-key architecture to avoid centralizing student data. Across the cohort, as a shared practice, no project reported collecting or retaining audio recordings of student interactions, and no project reported using student interaction data to train AI models. One faculty fellow discovered that configuring an AI tool with copyrighted course materials raised questions that existing policy did not answer. Rather than leaving the fellow to resolve the issue alone, the cohort worked collectively with institutional offices to obtain guidance, and what began as one project's obstacle became shared knowledge for the whole community.

On the urban-planning project, an intern implemented bias analytics, facial blurring, and metadata protections. On other projects, interns moved from a Translator role toward a Co-Pilot partnership as their judgment developed. The experience taught that research is a coordinated workflow with distributed responsibility. Trainees learn where AI can assist, where human expertise is required, and how to manage the boundary between the two. This is the practical meaning of learning by collaborating.

### Pilot Outcomes and Limits

The pilot produced 17 projects, 13 public presentations, five projects that reached classroom-ready status, and seven in active beta testing. Twelve of 15 surveyed fellows planned to continue their work into the summer or the next academic year [35]. Faculty also reported stronger confidence in evaluating AI tools and communicating about their disciplinary implications. Interns described growth in technical and professional judgment, including a more cautious understanding of AI's limits and the need for human oversight.

End-of-semester reviews provide an initial descriptive view of the polymathic competencies described in Section II.F. Faculty most often reported increased confidence in explaining AI's impact on their discipline and distinguishing meaningful tools from gimmicks. Interns most often reported gains in identifying ethical risks, understanding real-world AI use, communicating AI to non-technical collaborators, managing projects to deadlines, and professional prompt engineering (Figure 6).

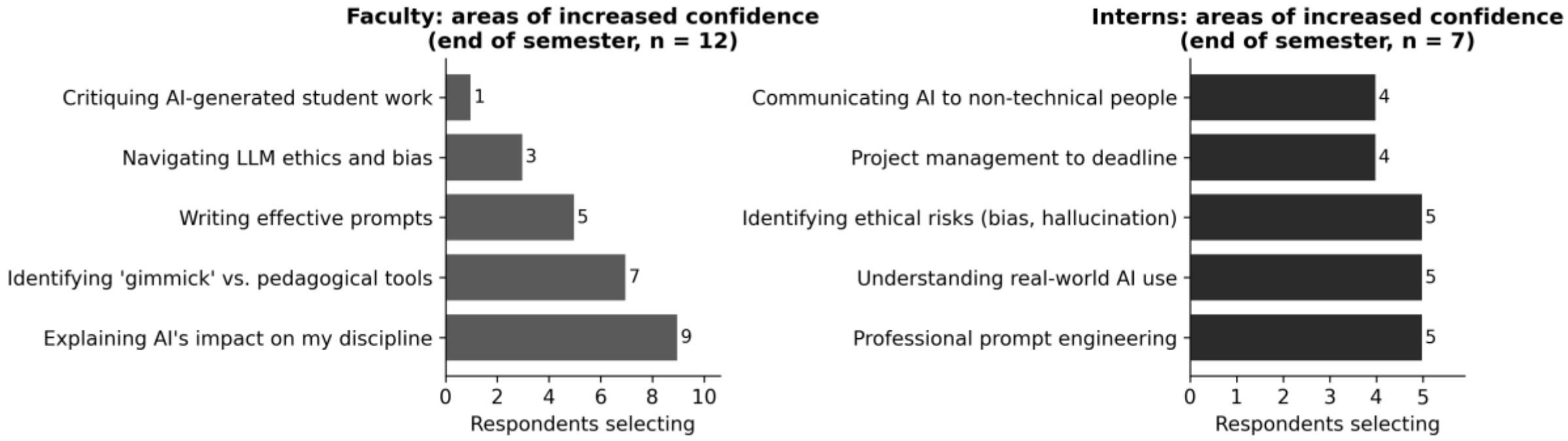

*Figure 6. Self-reported areas of increased confidence at the end of the semester. Faculty respondents (n = 12) and intern respondents (n = 7) selected multiple areas.*

These results are formative evidence for the mentoring model. They suggest that a shared studio can support technical learning together with translation, ethical reasoning, project management, and communication. The sample is self-selected and small, and the survey does not establish causal effects. Its value lies in showing how the model's intended capacities can be made visible and discussed within program practice.

These outcomes indicate that the model can organize a productive mentoring environment across fields. They also identify the resources required to sustain it: dedicated intern support, phased milestones, technical onboarding, policy guidance, data governance templates, funding for API and computation costs, and recognition for faculty participation. Participants also entered with different levels of technical confidence and benefited from support calibrated to their starting points. The evidence therefore informs both the intellectual model and its institutional conditions.

## IV. Conclusion

The aiX pilot of Everyone's Backyard (Section III) suggests that the deepest change AI brings to research mentorship is not only what mentors supervise but who mentors whom. Interns mentored faculty on AI tools, faculty gave interns the disciplinary context that rigorous design requires, and fellows from different schools mentored one another. When one fellow met an unresolved copyright question, the cohort sought institutional guidance together, and one project's obstacle became shared knowledge. Across these exchanges, what mentors contribute is judgment: asking worthwhile questions, evaluating evidence, interpreting uncertainty, making ethical decisions, and designing workflows that connect research to people and institutions.

Everyone's Backyard turns these observations into a design that other institutions can adapt. Its central move is to make mentoring reciprocal: each participant leads where their expertise is deepest and learns from others where it is not. Workflows, rather than isolated techniques, become the unit of mentorship; the convergence studio gives reciprocal mentoring a shared place to happen; and a community of practice carries methods and lessons across projects and cohorts. Statistical thinking, data ethics, and a commitment to AI for social good and society keep the work rigorous, responsible, and purposeful as it crosses disciplines.

The evidence remains early. Faculty and interns across 14 schools and departments built projects, formed a cross-disciplinary community, and reported gains in confidence and competence, but the pilot ran for one semester at one institution with self-selected participants and descriptive measures. Next steps include matched pre-post measures of expert and trainee capacities, following projects into sustained teaching and research use, and testing the model in settings with fewer resources, where the risk of widening gaps is greatest. The pilot also showed that the model depends on institutional support and on a willingness to treat unsettled questions as shared work. As AI continues to change how research is done, mentors will not prepare researchers by knowing every tool, but by building communities in which each person brings a discipline into the shared backyard and everyone is invited in. That is the work of research mentorship in the age of AI.

## Acknowledgments

The author thanks the faculty fellows and student interns of the aiX Faculty Fellows Program. This work was supported by a grant from Google to the Empire AI Consortium and by Columbia University's Office of the Vice Provost for Faculty Advancement, the Center for Teaching and Learning, the Data Science Institute, and Columbia University Information Technology.

## Conflict of Interest Statement

The author directs the aiX Faculty Fellows Program described in this chapter and declares no other competing financial or personal interests related to this contribution.

### Data Availability Statement

The quantitative findings reported here derive from internal program evaluation of the aiX Faculty Fellows Program (Spring 2026). De-identified summary tables and the analysis scripts used to generate the figures are available from the author on reasonable request. The findings are descriptive and should be interpreted in light of the self-selected samples, small review cohorts, retrospective confidence measures, and the absence of matched pre-post faculty responses.